\documentclass[prb,twocolumn,preprintnumbers,showpacs,amsmath,amssymb,superscriptaddress]{revtex4}
\usepackage{bm}
\usepackage{amsmath}
\usepackage[dvips]{graphicx}
\begin{document}
\bibliographystyle{apsrev}

\title{Weak antiferromagnetism and dimer order in quantum systems of coupled tetrahedra}

\author{Valeri N. Kotov}
\email{valeri.kotov@epfl.ch}
\affiliation{Institute of Theoretical Physics, Swiss Federal Institute of Technology (EPFL),\\
1015 Lausanne, Switzerland}
\author{Michael E. Zhitomirsky}
\affiliation{Commissariat \`a l'Energie Atomique, DSM/DRFMC/SPSMS, \\
17 avenue des Martyrs, 38054 Grenoble,  France
}
\author{Maged Elhajal}
\affiliation{Institute of Theoretical Physics, Swiss Federal Institute of Technology (EPFL),\\
1015 Lausanne, Switzerland}
\author{Fr\'ed\'eric Mila}
\affiliation{Institute of Theoretical Physics, Swiss Federal Institute of Technology (EPFL),\\
1015 Lausanne, Switzerland}

%\date{\today}
\begin{abstract}
We analyze the phases of an S=1/2 spin model on a lattice of coupled
tetrahedra. The presence of  both Heisenberg and antisymmetric, Dzyaloshinsky-Moriya
interactions  can lead to two types of symmetry-broken states: non-magnetic dimer order
and, unexpectedly, exotic  4 sub-lattice weak antiferromagnetic order - a
 state with  a generically small ordered moment and non-zero chirality.
External magnetic field also induces weak antiferromagnetism co-existing with
 strong dimer correlations in the ground state.
 These states are formed as a result
of broken Ising symmetries and exhibit a number of unusual properties.
 
\end{abstract}
\maketitle

\section{Introduction}
Some of the most challenging  and exciting problems in modern solid state
physics are related to the nature of symmetry-broken states and
the competition between different types of order  in  insulating and doped
antiferromagnets. \cite{Sachdev,Gregoire} There exist numerous materials,
ranging from
the high-temperature superconductors to molecular-based magnets, that 
provide continuous source of inspiration for this research.

In this paper we  study the types of order that
 can occur in quantum spin systems on lattices formed by coupled tetrahedra.
Perhaps the most well-known (and still not fully understood) model  of
this type 
is the 3D pyrochlore lattice, composed of corner-sharing
tetrahedra. \cite{Pyrochlore}
The purpose of this work  however is to look at a  class of models 
where the tetrahedra are connected weakly and in a more regular fashion,  almost in a 2D
square lattice-like arrangement. There are two main reasons for this.
First, such models allow for more reliable and complete theoretical treatment. 
Second,  
we have had in mind potential applications to 
 the $S=1/2$ material
Cu$_{2}$Te$_{2}$O$_{5}$Br$_{2}$ \cite{Peter} which is representative of such a geometry.
Rather unusually, in this material it has been observed that  
low-energy singlet excitations (measured in
 Raman spectroscopy) coexist with some kind of (possibly weak)
 magnetic order, the origin of which is still
 controversial. \cite{Peter,Gros}

We suggest that to understand the properties of the above and similar systems,
it is important to take into account antisymmetric,
Dzyaloshinsky-Moriya (DM) spin-spin interactions,
which are expected to be present in a tetrahedron on symmetry grounds. 
 Our main result, which is 
quite general and model dependent only in the details,
 is that under certain conditions the low-energy singlet dynamics
 can coexist with weak antiferromagnetism, induced by the DM interactions.
 It is  well known that weak magnetic moments can appear near a magnetic-paramagnetic
transition boundary, or in dimer (gapped) systems in the presence of
both DM interactions
 and external magnetic field. \cite{Shin}
In the present work  we describe a novel mechanism for weak 
antiferromagnetic order, which is induced by the DM interactions
even  without external fields. We show below that such an exotic
possibility exists in tetrahedral systems due to the degeneracy of the
 ground state on a single tetrahedron. 
The typical excitation signatures are different
from those of   conventional spin waves and 
various  nontrivial effects in an external magnetic field, such as magnetic
field induced order, are also present. 

We start with the following spin-1/2 Hamiltonian, which involves both Heisenberg
 and antisymmetric exchanges:
%\vspace{-0.2cm}
\begin{equation}
\hat{{\cal H}} = \sum_{{\bf i,j}}J_{{\bf i,j}} {\bf S}_{\bf i}.{\bf S}_{\bf j}
 + \sum_{{\bf i,j}} {\bf D}_{{\bf i,j}}.({\bf S}_{\bf i} \times {\bf S}_{\bf j}),
\label{ham}
\end{equation}
where the couplings $J_{{\bf i,j}}$ are distributed as shown in Fig.~1(b).
\begin{figure}[ht]
\centering
\includegraphics[height=115pt, keepaspectratio=true]{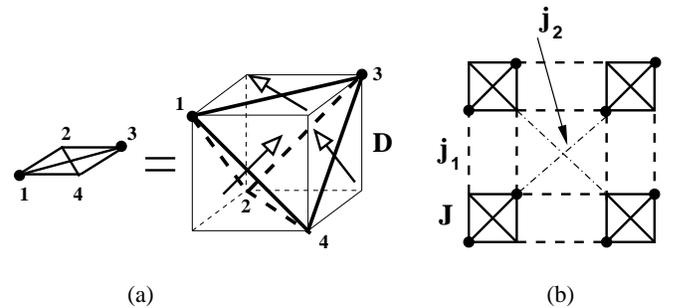}
\caption{ (a) A tetrahedron with DM vectors represented by arrows (three of the
 six shown). (b)
Two-dimensional lattice of coupled tetrahedra.} 
\label{Fig1} 
\end{figure}
\noindent
The tetrahedra are represented as plaquettes in
Fig.~1(a), and are assumed to be coupled weakly, i.e. $0 \leq j_{1},j_{2} \ll J$,
where $J$ is the exchange inside one tetrahedron. All couplings
are antiferromagnetic. We analyze the model in two dimensions but the inclusion
of three-dimensional couplings is straightforward and does not affect our main
results.  
 For simplicity of analysis and notation we set for now $j_{2}=0$, and
the effect of $j_{2}$ on our results will be discussed later.
The DM couplings have relativistic origin, \cite{Dz,Moriya} 
and are expected to be weak, usually at most several percent of
the Heisenberg exchange, $|{\bf D}_{{\bf i,j}}| \ll J$. The distribution of the
couplings ${\bf D}_{{\bf i,j}}$ is very lattice-specific, as the presence of the
DM interaction on a particular bond is determined by the symmetry of the environment.
We will make the assumption that the ${\bf D}_{{\bf i,j}}$'s are present
only on the tetrahedra  (where the Heisenberg exchange is also dominant),
and are non-zero on every bond within one tetrahedron.   
%\cite{remark1}.
 The distribution of the six vectors  ${\bf D}_{{\bf i,j}}$,
written with respect to the principal axes of the cube in Fig.1(a), and consistent with
the tetrahedral group is:  
${\bf D}_{13}\! =\! \frac{D}{\sqrt{2}}(-1,1,0)$, ${\bf D}_{24}\! =\! \frac{D}{\sqrt{2}}(-1,-1,0)$,
${\bf D}_{43}\! =\! \frac{D}{\sqrt{2}}(0,-1,1)$, ${\bf D}_{12}\! =\! \frac{D}{\sqrt{2}}(0,-1,-1)$,
${\bf D}_{14}\! =\! \frac{D}{\sqrt{2}}(1,0,1)$, ${\bf D}_{23}\! =\! \frac{D}{\sqrt{2}}(1,0,-1)$.
 Here $D$ is the magnitude of the  DM vectors.

The rest of the paper is organized as follows. In Section II
we derive the effective Hamiltonian
in the weak-coupling limit. The quantum phases of the model are then
 analyzed in Sections III and IV. 
Section V contains our conclusions.

\section{Effective Hamiltonian in the presence of DM interactions}

First, we describe the states of one tetrahedron, Fig.~1(a). For $D=0$
the ground state is twofold  degenerate (with energy $E_{0}=-3J/2$)
 and is in the singlet sector 
$S^{tot}=0$. The two ground states are: 
$|s_1\rangle=\frac{1}{\sqrt{3}}\{[1,2][3,4]+[2,3][4,1]\}$, 
$|s_2\rangle=\{[1,2][3,4]-[2,3][4,1]\}$,
where $[k,l]$ denotes
a singlet formed by  the nearest-neighbor spins $k$ and $l$.
The DM interactions break the continuous spin rotational
invariance, leading to an 
admixture of triplets to the ground state. \cite{remark2}  
Introducing the three excited (energy 
$E_{1}=-J/2$)  $S=1$ triplet states $p_{\mu},q_{\mu},t_{\mu}$, $\mu =x,y,z$,
in the notation of Ref.~\onlinecite{Misha},
 we obtain the two  new ground states, in the limit $D/J \ll 1$:
%\vspace{-0.2cm}
\begin{eqnarray}
\label{DMsinglets}
|\Phi \rangle & = & |s_1 \rangle +
\frac{3iD}{2\sqrt{6}J}\left[|p_x\rangle -|p_y\rangle +|q_x\rangle
 +|q_y\rangle \right] \label{wave}
 \\
|\Psi \rangle & = & |s_2 \rangle +
\frac{iD}{2\sqrt{2}J}\left[|p_x\rangle +|p_y\rangle +|q_x\rangle
 -|q_y\rangle \right] + i\frac{D}{J} |t_z\rangle. \nonumber
\end{eqnarray}
These  states remain degenerate with 
$E_{0}^{DM}=-3J/2- 3D^{2}/(2J)$, and their  
wave-functions transform according to the $e_g$ irreducible representation
of the tetrahedral point group. 
The diagonal matrix elements of on-site spins between the modified ground states (\ref{wave})
are zero, whereas the off-diagonal matrix elements acquire finite {\it imaginary}
values $\langle\Psi|S_{\alpha,n}|\Phi\rangle = \pm iD/\sqrt{6}J$. 

The low-energy sector of the Hilbert space has dimensionality $2^{N_{\rm tet}}$, where
$N_{\rm tet}$ is the number of spin tetrahedra.
The remaining degeneracy is lifted by the inter-tetrahedral couplings, and we proceed 
 to analyze the nature of the new ground  state.
 For this purpose we introduce
a pseudo-spin ${\bf T}=1/2$ representation, so that
$T_{z} = 1/2$  corresponds to  $|\Phi  \rangle$
and  $T_{z} = -1/2$ corresponds to  $|\Psi \rangle$. We then obtain the effective
Hamiltonian in the ground state subspace:
%\vspace{-.1cm}
\begin{eqnarray}
\hat{{\cal H}}_{\mbox{eff}}&=& - \sum_{\langle {\bf i,j}\rangle}
\Bigl[ \Omega_{x} T_{x,{\bf i}}T_{x,{\bf j}} +
\Omega_{z} T_{z,{\bf i}} T_{z,{\bf j}}
  + \Omega_{y} T_{y,{\bf i}} T_{y,{\bf j}} \nonumber \\
 &+ &\Omega_{xz}^{({\bf i},{\bf j})}
 (T_{z,{\bf i}}T_{x,{\bf j}} \!+ \!T_{x,{\bf i}}T_{z,{\bf j}})\Bigr]
- h  \sum_{\bf i} T_{z,{\bf i}}  
\label{pseudoham}
\end{eqnarray}
Now the site indexes ${\bf i},{\bf j}$ refer to the positions of the
tetrahedra, and the summation is over nearest neighbors on a square lattice. 
To lowest order in  $j_{1},D$ 
 the couplings in the
different pseudospin directions are explicitly:
%\vspace{-0.2cm}
\begin{eqnarray}
\label{ef1}
\Omega_{x} = \frac{j_{1}^{2}}{4J},\ \Omega_{y}=j_{1}\frac{4 D^{2}}{3 J^{2}},
 \ \Omega_{z}= \frac{\Omega_{x}}{3},
 \ h= \frac{2}{3}\Omega_{x}.
\label{couplings}
\end{eqnarray}
The mixed coupling depends on the bond direction
$\Omega_{xz}^{({\bf i},{\bf j})}=\Omega_{x} e^{i{\bf Q}.({\bf i}-{\bf j})}/\sqrt{3},
\ {\bf Q}=(\pi,0).$
\vspace{0.2cm} 
\begin{figure}[ht]
\centering
\includegraphics[height=155pt,keepaspectratio=true]{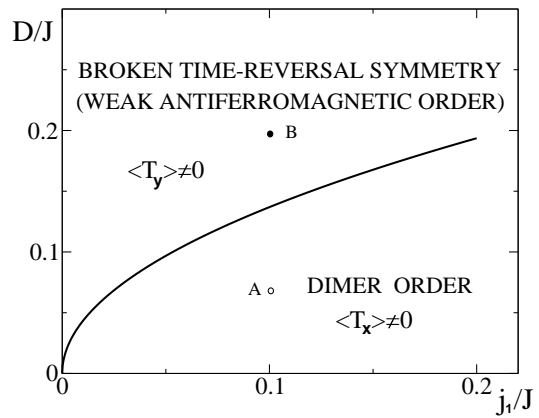}
\caption{Phase diagram of the pseudospin Hamiltonian at zero temperature.
}
\label{Fig2}
\end{figure}

%We proceed to analyze the ground states of $\hat{{\cal H}}_{\mbox{eff}}$, Eq.(\ref{pseudoham}).
The effective Hamiltonian describes an anisotropic ferromagnet in an effective magnetic field $h$
applied along the ``hard'' $z$-axis. If the field $h$ does not exceed a critical
value $h_c$ (which we find to be the physical case, corresponding to
(\ref{couplings})),
 the system breaks the Ising symmetry along the soft $x$ or  $y$ axes.
The selected direction in the $xy$-plane depends on $\Omega_x/\Omega_y$.
The first-order transition line 
$\Omega_{x}=\Omega_{y}$, on mean-field level, separates regions with
$\langle T_{x, {\bf i}}\rangle \neq 0, \langle T_{y, {\bf i}} \rangle=0$
(for $\Omega_{x}>\Omega_{y}$),
 and $\langle T_{y, {\bf i}}\rangle \neq 0, \langle T_{x, {\bf i}} \rangle=0$
(for $\Omega_{y}>\Omega_{x}$).
The two regions are shown in Fig.~2 as a function of the microscopic parameters.  
Since $h\neq0$, throughout the phase diagram 
$\langle T_{z, {\bf i}}\rangle \neq 0$. We now discuss in more
detail the nature of symmetry breaking in the two phases.
%\vspace{0.2cm}
\begin{figure}[ht]
\centering
\includegraphics[height=105pt,keepaspectratio=true]{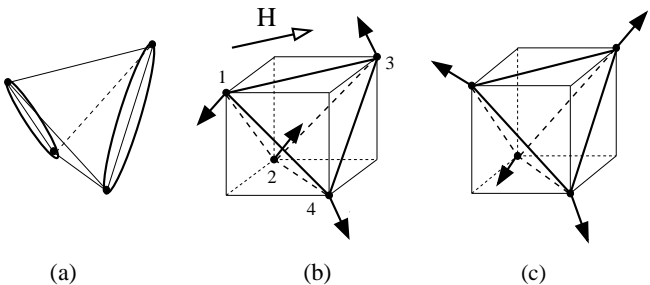}
\caption{(a) Dimer state, $\langle T_{x, {\bf i}} \rangle \neq 0$.
 (b) Magnetic moments in an external magnetic field along
the 1-3 bond in the dimer phase from (a)
(tilt of spins in the field direction not shown). (c) Spin arrangement in the 
phase  $\langle T_{y, {\bf i}} \rangle \neq 0$, at zero  field.
}
\label{Fig3}
\end{figure}

\section{Dimer phase and magnetic field induced order}
 The
phase with $\langle T_{x,{\bf i}}\rangle \neq 0$ corresponds to dimer
order. The existence of such a phase for the case $D=0$ was pointed
out in Ref.~\onlinecite{Kotov}. Physically, this means 
that the  ground state of the system is the linear combination 
$\frac{\sqrt{3}}{2}|\Phi\rangle\pm\frac{1}{2}|\Psi\rangle$, on every site,  
 which is equivalent to 
 $\langle T_{x, {\bf i}} \rangle=\pm\sqrt{3}/4, \langle T_{z, {\bf i}}\rangle=1/4$
 (these values correspond to (\ref{couplings})).
The ground state is a real combination of $|\Phi\rangle$ and $|\Psi\rangle$,
and therefore the expectation values of local spins vanish, and  only dimer order 
is present 
in this case, as
 shown in Fig.~3(a).
Since the $\Omega_{x}$ coupling
in (\ref{pseudoham}) has ferromagnetic sign, the dimer pattern repeats
itself on all tetrahedra.  
Magneto-elastic couplings could also contribute to the dimerization
tendency as discussed for the pyrochlore case, \cite{Yamashita} but in our
scenario dimerization occurs spontaneously  and is  due to purely Heisenberg exchanges.

The dimer ordered state  breaks the discrete, Ising symmetry $T_{x, {\bf i}}
\rightarrow -T_{x, {\bf i}}$ of (\ref{pseudoham}), and thus the excitation
spectrum $\omega({\bf k})$ has a gap
$\Delta = \omega({\bf k=0})= 1.43 \Omega_{x}$ (for $D=0$). At finite temperature $T$,
the ordered state exists below the critical temperature $T_{c}=0.92 \Omega_{x}$.
 We have used  the linear spin wave expansion and the mean-field equations
to calculate these quantities.
\vspace{0.3cm}
\begin{figure}[ht]
\centering
\includegraphics[height=170pt,keepaspectratio=true]{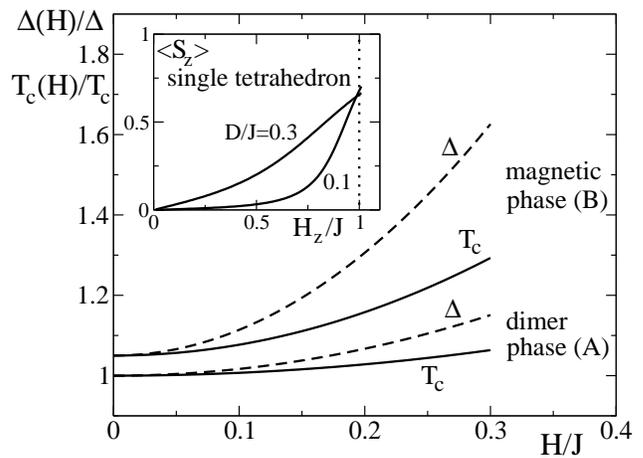}
\caption{Dependence of the gap (dashed lines) and $T_{c}$ (solid lines) on
magnetic field (applied as in Fig.~3(b)) at the points A and B of Fig.2. 
Two upper curves shifted by 0.05 for clarity. Inset: Magnetic moment
$\langle S^{tot}_z \rangle$ in magnetic field in the  z direction for a single
tetrahedron (results identical in x,y directions). For $D=0$, $\langle S^{tot}_z \rangle=0,
H_{z}/J < 1$.}
\label{Fig4}
\end{figure}

Due to the presence of triplets in the ground state (\ref{DMsinglets}), a (real) external
uniform magnetic field $-g\mu_{B}{\bf H}\cdot{\bf S}^{tot}$ also leads to non-trivial effects.
On a single tetrahedron ($j_{1}=0$) the field produces a finite magnetization
in the field direction, shown in the inset of Fig.~4. For small fields $H/J \ll 1$,
the contribution is linear and proportional to: $(D/J)^{2}(H/J)$. We find that
the  magnetization
is only weakly dependent on the field direction.
For $j_{1}\neq0$, the corrections to the effective
Hamiltonian (\ref{pseudoham}) have to be derived, and we find them to be 
strongly dependent on the field direction. For example for
a weak field in the x-y direction ${\bf H}=\frac{H}{\sqrt{2}}(1,1,0)$, 
Eq.~(\ref{DMsinglets}) changes to: 
%\vspace{-0.2cm}
\begin{eqnarray}
|\Phi \rangle_{H} & = & |\Phi \rangle 
-\frac{\sqrt{3}DH}{2J^{2}}|p_z\rangle 
 \\
|\Psi \rangle_{H} & = & |\Psi \rangle - 
\frac{DH}{2J^{2}}[|q_z\rangle +
\sqrt{2}(|t_y\rangle -|t_x\rangle)].
 \nonumber
\label{Fieldsinglets}
\end{eqnarray}
The energy difference between these states leads to
a modification of the "effective field" $h \rightarrow h - D^{2}H^{2}/(2J^{3})$
in Eq.(\ref{pseudoham}). 
Even though several additional terms are generated in $\hat{{\cal H}}_{\rm eff}$,
for $j_{1},D,H \ll J$  the above effect is dominant. We have recalculated
the gap $\Delta(H)$ and the critical temperature $T_{c}(H)$ in the presence
of the field and the results are summarized in Fig.~4, where we have plotted
these quantities relative to their zero field values. The gap generally shows 
 a stronger field dependence. We note that the increase as a function of field
is not universal, and in fact if a field is applied in the z direction
${\bf H}=(0,0,H)$,
then  $\Delta(H)$ and $T_{c}(H)$ would decrease at almost the same rate.   

An applied magnetic field also generates staggered antiferromagnetic
moments in the plane
perpendicular to the field. The corresponding  pattern is shown in Fig.~3(b),
where $\langle {\bf S}_3\rangle = -\langle {\bf S}_4\rangle$, 
$\langle {\bf S}_1\rangle = -\langle {\bf S}_2\rangle$,
and the spins on sites $2$ and
$4$ point along the diagonals of the cube (towards and out of the cube's center,
respectively). The value of the magnetic moments in this pattern,
 repeated on all tetrahedra,  is
\begin{equation}
|\langle {\bf S}_n \rangle| = \frac{DH}{J^{2}}\langle T_{x, {\bf i}}\rangle =
 \frac{\sqrt{3}DH}{4J^{2}}, \ \ \ n=1,2,3,4.
\end{equation}
Field induced antiferromagnetic ordering is a general feature
of singlet systems with  DM interactions.

\section{Broken time-reversal symmetry state with weak antiferromagnetic order}
 
The phase with 
$\langle T_{y, {\bf i}} \rangle  \neq 0$ in Fig.~2 corresponds
to a ground state which is a complex linear combination on a single tetrahedron 
 $\alpha|\Phi \rangle \pm i \beta |\Psi \rangle$, where $\alpha, \beta$ are real
coefficients.
This combination is ferromagnetically repeated on every tetrahedron.
 Depending  on the values of the microscopic parameters,
 $\sqrt{3}/4 \leq |\langle T_{y, {\bf i}} \rangle|
\leq 1/2$. 
The resulting state resembles to a large extent
a spin-liquid state with a broken time-reversal symmetry proposed
some time ago \cite{Wen} and characterized by   finite scalar chirality
$\chi = \langle {\bf S}_1\cdot({\bf S}_2 \times {\bf S}_3) \rangle \neq 0$. 
In our case such a state appears, however, due to the anisotropic DM
interaction in a system without spin-rotational invariance,
 and we have $\chi \sim \langle T_{y, {\bf i}} \rangle$. Therefore,
a broken time-reversal symmetry immediately induces finite antiferromagnetic
moments
with magnitude: 
%\vspace{-0.1cm}
\begin{equation} |\langle {\bf S}_n \rangle| = \frac{D}{J}
\sqrt{2} \langle T_{y, {\bf i}} \rangle.
\end{equation}
The spins are at an angle $\phi = 109.47^\circ$ with respects to each other,
and form a four-sublattice antiferromagnetic structure, as shown in Fig.~3(c). 
On each tetrahedron the total magnetic moment
 $\sum_{n}\langle {\bf S}_n \rangle=0$. 
 
Similarly to the dimer phase, the broken Ising symmetry leads to 
an excitation gap and a finite Ising transition temperature, 
whose values, for dominant DM interactions are respectively
$\Delta\simeq \Omega_y$ and $T_c\simeq  \Omega_y$.
Thus, we find a quite unusual situation: 
the specific heat exhibits a large
anomaly at the transition temperature, where a macroscopic 
part of the low-temperature entropy $N_{\rm tet}\ln 2$ freezes out.
However, if looked at in neutron experiments such a transition
is characterized by development  of small antiferromagnetic moments below $T_c$.
This exotic behavior is due to the fact that the transition itself is driven by
low-energy singlet degrees of freedom.
In applied magnetic field ${\bf H}=\frac{H}{\sqrt{2}}(1,1,0)$, the gap and 
$T_c$ in the broken time-reversal symmetry state scale as shown in Fig.~4, upper
set of curves.  
We find that generally the field dependence is stronger than in the
dimer phase, which is mainly due to the presence of the DM scale $D$ in the
 two observables without a field. 

%\begin{figure}[ht]
%\centering
%\includegraphics[height=140pt,keepaspectratio=true]{graph3.eps}
%\caption{Spin arrangement in the phase $\langle T_{y, {\bf i}}\rangle \neq 0$}
%\label{Fig5}
%\end{figure}

We note that $\Omega_{y}$ depends linearly on $j_{1}$ while $\Omega_{x}$
is quadratic (\ref{couplings}). Thus, even though the $T_{y}$ (magnetic)  and the $T_{x}$ (dimer)
 orders compete with each
other, at sufficiently small $j_{1}$ the $T_{y}$ order is expected to dominate
(although only in asymptotic  sense for the present model). 
The tendency towards $T_{y}$ order is further enhanced by the presence of $j_{2}\neq0$
(see Fig.~1(b)). This coupling frustrates the $T_{x}$ component while
strengthening  $T_{y}$, which results in a shift of the phase boundary in Fig.~2 downwards.
 If we also consider a situation when $j_{2}$
 is the dominant inter-tetrahedral exchange, $J \gg j_{2} \gg j_{1}$, and set $j_{1}=0$,
then we find that  $T_{x}$ couplings  are generated only in fourth order,
$\Omega_{x} \sim j_{2}^{4}/J^{3}$. This again means that the $T_{y}$ order
can occur under much easier conditions.
Nonetheless, only on lattices that have the tendency to produce true
spin liquid ground states do we expect the $T_{y}$ order to occur
spontaneously. Good candidates appear to be the pyrochlore lattice
\cite{Pyrochlore} as well as the kagom\'e lattice. \cite{Mila} DM
induced ordering has been discussed in the latter case from quasiclassical
perspective. \cite{Maged} 

\section{Discussion and Conclusions}

Our calculations have been performed for  ideal tetrahedra 
while  weak distortions (such as asymmetry in the Heisenberg exchanges,
or deviations of the DM vectors from the tetrahedral symmetry) would introduce
a small gap on a single tetrahedron.
  However, as long as 
 $j_{1}, j_{2}$ are sufficiently large, relative to this gap, 
the  basic characteristics of the phases  described above  will remain unchanged.
 The DM order may deviate from  Fig.~3(c), while preserving the 
property  $\sum_{n}\langle {\bf S}_n \rangle=0$. 

The physics described in the present work occurs on the
 small energy scales $\Omega_{x}, \Omega_{y} \ll J$,
while upon increasing 
the inter-tetrahedral couplings  to
$j_{1},j_{2} \sim J$, a transition to a more conventional
N\'eel phase is expected to  take place. \cite{Gros, Brenig} 
The exotic phases we have found are certainly  stable as long as 
the system is sufficiently far from the N\'eel order, i.e. the triplet gap 
is non-zero and all the relevant dynamics is governed by the
 singlet sector (or the mixed states Eq.~(2) in the presence of
DM interactions), and therefore can be described by the
 effective Hamiltonian Eq.~(3). We have made sure that this condition
is fulfilled by using the weak-coupling analysis,
 although our results are expected
 to be qualitatively valid also for moderately large $j_{1},j_{2}$ subject, as already mentioned,
 to the condition that the triplet gap is sizable enough.
 Thus, on general grounds we expect  a transition as a function of
 $j_{1}$ or $j _{2}$  between phases with exotic (weak magnetic) order,
discussed in this work,  and 
 the N\'eel ordered state, although the exact determination of the transition
 boundary  is a difficult problem and has not been the subject of this work.
 Instead, we have concentrated on the ``universal'' properties of the
 DM induced exotic phases  which  are clearly present in the weak-coupling limit
and accurately described by a Hamiltonian of the form Eq.~(3).

The states we have found at low energies have quite unusual characteristics,
which we will now summarize. (i) Due to the presence of the DM interactions
the spin-rotational invariance is broken and the ground state develops a small magnetic moment
in a field, with     
 the  uniform magnetization  showing a characteristic linear behavior in
weak magnetic fields (Fig.~4(inset)).
(ii) Either dimerized or 4-sublattice, weakly antiferromagnetic states
emerge as the ground state of the system. Both exhibit gaps in
 the spectrum since only Ising symmetries are broken. 
(iii) The dominant components in the wave functions Eq.(\ref{DMsinglets}) are singlets,
thus a large spectral weight S=0 transitions at low energy
(equal to the gap $\sim \Omega_{x}, \Omega_{y}$) are present in the Raman spectrum.
(iv) Both the gap and $T_{c}$ depend on magnetic field with a strong
directional dependence.
(v) The dominant S=1 excitation has a large gap of order
 $J \gg \Omega_{x}, \Omega_{y}$.

We now comment on possible applications of our theory to 
  Cu$_{2}$Te$_{2}$O$_{5}$Br$_{2}$. The properties of this material
 are indeed consistent with signatures (i-iv) above, in particular 
the compound exhibits: \cite{Peter,Gros} 
(a.) a weak magnetization in a field, (b.) a sharp zero-field, low-energy
 Raman peak ("singlet gap" $\Delta \approx 24 K$), disappearing at $T_{c}\approx 11K$,
(c.) a magnetic field dependence (increase) of both $\Delta$ and $T_{c}$.
Experimentally  the 
most important 
unanswered 
question  remains  whether the ground state below $T_{c}$ 
is magnetically ordered or not, with recent NMR experiments  pointing to 
some kind of magnetic order. \cite{Claude}  This would
suggest, in our scenario, that the ground state is of the type shown in Fig.~3(c). The crucial
test for our ideas, and thus our prediction, would be the observation of a large triplet gap
(larger than the "singlet gap" of 24K) in the
neutron spectrum (point (v) above). 
The microscopic values of the exchanges  are still rather
controversial \cite{Johnsson} and consequently we have not attempted detailed
 fits but have used, phenomenologically, the model of Fig.~1. 
 We emphasize that in our model
 a weak magnetic moment is a generic
feature of the $\langle T_{y, {\bf i}} \rangle \neq 0$ phase
(and the dimer phase in magnetic field),
 and does not require a fine-tuning of parameters
 to achieve proximity to a magnetic-paramagnetic boundary, as in the purely
 Heisenberg case. \cite{Gros}

Finally, on purely fundamental level, we  have
found an unconventional mechanism for {\it weak antiferromagnetism}. 
Our results are  to be contrasted with the usual effect the  DM interaction
produces, \cite{Dz,Moriya} namely weak ferromagnetism in an otherwise
 antiferromagnetic system. 
In the present work we have shown how weak antiferromagnetic order can emerge from a
 singlet background. 

\begin{acknowledgments}
We are grateful to C. Lhuillier, C. Lacroix, O. Sushkov, and C. Berthier
for stimulating discussions. This work was supported by MaNEP and  the Swiss National Fund.
\end{acknowledgments}


\begin{thebibliography}{99}
\bibitem{remark2} The S=2 excited states do not mix with the ground state.
\end{thebibliography}
\end{document}